\documentclass[%
reprint,
superscriptaddress,
footinbib,
longbibliography,
noeprint,
amsmath,amssymb,
aps,
]{revtex4-1}
\usepackage{amsthm}

\usepackage{graphicx}
\usepackage{physics}
\usepackage{xcolor}
\usepackage[autostyle, english = american]{csquotes}
\usepackage[ruled,vlined, linesnumbered]{algorithm2e}

\usepackage{bm}  
\renewcommand{\vec}[1]{\boldsymbol{\mathbf{#1}}}

\MakeOuterQuote{"}
\usepackage[caption=false,justification=centerlast]{subfig}
\definecolor{darkGreen}{RGB}{0,110,0}
\definecolor{darkBlue}{RGB}{0,0,130}
\usepackage[colorlinks,citecolor=darkGreen,linkcolor=darkBlue,urlcolor=blue,hyperindex]{hyperref}

\def\equationautorefname~#1\null{Eq. (#1)\null}

\newcommand{\appref}[1]{\hyperref[#1]{App.~\ref*{#1}}}

\newcommand{\comment}[1]{}

\begin{document}
\title{Strong-randomness renormalization groups}
\date{\today}
\author{David A. Huse}
\affiliation{Department of Physics, Princeton University, Princeton, New Jersey 08544, USA}
\begin{abstract}
This is a very brief review article, written for a book (in preparation) in memory of Michael E. Fisher and to celebrate 50+ years since the Wilson-Fisher renormalization group.   Strong-randomness renormalization groups were first developed to treat various quantum critical ground states, especially in one-dimensional systems.  After briefly reviewing some of the earlier work with these methods, the recent application of this approach to the many-body localization (MBL) phase transition is reviewed.
\end{abstract}

\maketitle

\section{Introduction}

Renormalization group approaches to systems in critical states have been developed for a wide variety of different types of systems.  As a consequence of this variety, there are many very different types of renormalization group (RG) calculations.  What do they all have in common?  Almost all RGs coarse-grain a system of many degrees of freedom by some type of ``integrating out'' of the highest-energy or in some sense ``stiffest'' degrees of freedom.  Good analytical control of the RG usually requires some level of simplicity of the couplings between the degrees of freedom that are integrated out and the remaining degrees of freedom.  

For many translationally-invariant systems the RG is implemented in momentum space, as in the seminal Wilson-Fisher paper~\cite{wilsonfisher,mefrgrmp}.  In many cases, the highest-energy (or stiffest) degrees of freedom are the remaining modes with the highest momenta, and these are integrated out, thus decreasing the ultraviolet momentum cutoff.  For systems of many degenerate fermions ``highest momenta'' may be replaced with momenta farthest in energy from the Fermi energy (see, e.g., Ref.~\cite{shankar}).  

Some other renormalization groups work instead in real space, one notable example being the Kosterlitz-Thouless RG for two-dimensional superfluids, which can be formulated with a real-space cutoff on the distance between a vortex and an antivortex~\cite{kt,kosterlitz}.  Vortex-antivortex pairs at this cutoff distance are 
integrated out, thus increasing the cutoff distance.  The good analytic control of the Kosterlitz-Thouless RG is due to the fixed line governing the superfluid phase and its critical point being at zero density of vortices and antivortices, so the pairs that are being integrated out are dilute when we are near the fixed line.   A similar real-space RG for some one-dimensional spin systems, where one integrates out pairs of domain walls at the cutoff distance, was developed earlier, in Ref.~\cite{ayh}.

For systems with strong quenched spatial randomness, the highest-energy (or stiffest) degrees of freedom are typically at some particular location in real space where the couplings are the strongest.  This naturally leads to a real-space RG, as is discussed below, and as has been reviewed thoroughly by Iglói and Monthus \cite{im, im2}, and more briefly by Refael and Altman \cite{ar}.  Such strong-randomness renormalization groups are the focus of this very brief review, which begins with some discussion of their use for quantum critical ground states, and finishes with more recent generalizations that address the many-body localization (MBL) phase transition that can occur in highly excited quantum states.

\section{random-singlet ground state}

The first strong-randomness RG was developed by Ma, Dasgupta and Hu (1979) for the low temperature properties of random quantum antiferromagnets~\cite{mdh,dasgupta,bhattlee,dsf94}. The system they considered is an antiferromagnetic Heisenberg spin-1/2 chain, with quenched random nearest-neighbor spin interactions. The Hamiltonian is:
\begin{equation}
H = \sum_n J_n\vec\sigma_n\cdot\vec\sigma_{n+1}~,
\end{equation}
where the couplings $J_n>0$ are drawn from a probability distribution $P(\log J)$ that is broad, and $\vec\sigma_n$ is the vector of Pauli operators for the spin-1/2 at site $n$.  The ground state of this spin chain is a so-called random-singlet state~\cite{mdh,dasgupta,bhattlee,dsf94}. This random-singlet ground state consists, to first approximation, of pairs of spins that are in their total spin zero singlet state, with the two spins in each such pair being at various distances along the chain determined by local details of that sample's Hamiltonian.

The strong-randomness renormalization group treatment of this random-singlet ground state proceeds as follows:  Each coarse-graining step of the RG ``integrates out'' the two spins, $n$ and $(n+1)$, that are coupled by the strongest remaining renormalized coupling $J_n$.  The order in which the spins are integrated out is dictated by energy, so proceeds in a sequence that is specific to each particular sample.  Since these pairs of strongly-coupled spins are each located at some position in real space, this RG is often called a real-space RG.  But really it more fundamentally works in ``energy space''.  The two spins to be integrated out are also coupled to their other neighbors via the couplings $J_{n-1}$ and $J_{n+1}$, which are both weaker than $J_n$.  The small parameters that allow good analytic control of this RG are the ratios $J_{n-1}/J_n$ and $J_{n+1}/J_n$.  When the distribution $P(\log{J})$ is broad, this makes these ratios typically small, so this RG approach is a controlled approximation.  Assuming that these two ratios are small, those weaker couplings can be treated in low-order perturbation theory.  

Thus we have the three terms in the local Hamiltonian:
\begin{equation}
H = ...+ J_{n-1}\vec\sigma_{n-1}\cdot\vec\sigma_{n}+ J_n\vec\sigma_n\cdot\vec\sigma_{n+1}
+J_{n+1}\vec\sigma_{n+1}\cdot\vec\sigma_{n+2} +...,
\end{equation}
with the middle term the strongest.  The lowest order approximation to the ground state is: spins $n$ and $(n+1)$ are in their total spin zero singlet state.  If we stop at that very lowest perturbative order, the chain is cut, so this does not address what are the ground-state correlations between, for example, spins $(n-1)$ and $(n+2)$.  The leading term that produces an interaction across the two spins that we are ``integrating out'' is produced at second order in perturbation theory in the weaker couplings, resulting in the renormalized Hamiltonian:
\begin{equation}
H' = ...+\delta E_0+ J'\vec\sigma_{n-1}\cdot\vec\sigma_{n+2}+...,
\end{equation}
with renormalized coupling $J'=J_{n-1}J_{n+1}/(2 J_n)+...$ as well as a contribution $\delta E_0=-3J_n+...$ to the ground state energy.  The ...'s refer to higher-order perturbative effects, which are asymptotically RG-irrelevant.  So this RG step puts spins $n$ and $(n+1)$ in the local ground state, thus removing those degrees of freedom, and introduces a new renormalized coupling between spins $(n-1)$ and $(n+2)$, producing a renormalized spin chain that is two spins shorter in length.

As the initial authors understood~\cite{mdh,dasgupta,bhattlee}, and as was analysed in much more detail by Daniel Fisher~\cite{dsf94}, this approach results in a {\it functional} renormalization group for the probability distribution $P(\log J)$ of the nearest-neighbor couplings.  Under the RG flow, this distribution broadens without limit due to the very weak new renormalized couplings that are produced.  This produces a RG flow to an {\it infinite-randomness fixed point}, where the approximations $J_{n\pm 1}\ll J_n$ used in the RG become asymptotically exact, so this RG gives the correct low-energy description of this system.  Because the three couplings that entered in setting the renormalized coupling $J'$ are all removed from the renormalized Hamiltonian $H'$, no correlations between different couplings $J_m$ are generated by the leading-order RG.  Thus although one could have a joint probability distribution $P(\{J_m\})$ with short-range correlations between the couplings on different bonds, such correlations are RG-irrelevant: the RG flow reduces these correlations and the fixed point distribution is asymptotically uncorrelated, which facilitates the analysis \cite{dsf94}.

This random-singlet ground state is a quantum critical state. Any spin-1/2 chain of this form, with random nearest-neighbor-only antiferromagnetic couplings drawn from a continuous joint probability distribution with only short-range correlations in the couplings, has a ground state governed by this same infinite-randomness fixed point.  Asymptotically, when the remaining spins are at typical distance $r$ bare lattice spacings, the typical renormalized couplings $J$ scale as: $-\log{J}\sim r^{1/2}$ for large $r$.  Thus there is an exponential dynamical scaling (dynamical critical exponent $z\rightarrow\infty$), with the energy scales decreasing as the exponential of a power of the length scale.  Such exponential dynamical scaling is typical of infinite-randomness quantum critical points.  The equal-time spin-spin correlations in this ground state are very broadly distributed, with the mean correlation falling off with distance as $\sim r^{-2}$ due to rare strongly-correlated spins, while the typical correlations fall off with distance exponentially in $r^{1/2}$, as do the renormalized couplings \cite{dsf94}.

\section{other infinite-randomness fixed points}

There are (infinitely) many other one-dimensional models, both quantum and classical, whose asymptotic ground state and/or low-frequency properties can be systematically and correctly obtained with a strong-randomness RG approach \cite{im,im2,ar}.  One notable example is the quantum critical point of the transverse-field Ising spin chain, whose infinite-randomness fixed point is very closely related to that of the random-singlet state \cite{dsf92,dsf95}.  An infinite series of discretely different infinite-randomness fixed points was also found.  These govern the quantum critical ground states of certain spin chains with larger spin $S$, as well as chains of interacting non-Abelian anyons \cite{im,im2,dh,frlt}.

For many of these one-dimensional models, the strong-randomness RG can be solved analytically.  When the method is generalized to systems in more than one dimension ($d>1$), parts of the calculations must instead be implemented numerically. It is found that, unlike in one dimension, for $d>1$ and most models the RG flow near infinite randomness is instead towards weaker randomness, so such systems are not governed by infinite-randomness fixed points.  The principal exception to this is the quantum critical point of the random transverse-field Ising model, which appears to remain governed by infinite-randomness fixed points for all dimensions $d$ \cite{im,im2}.

\section{many-body localization}

One recent development in the use of strong-randomness renormalization groups is in the study of many-body localization (MBL).  Some of this RG work was reviewed in Refs.~\cite{im2,ar}, so here I will focus on the more recent work.  

Many-body localization (MBL) is Anderson localization for many interacting particles or spins in highly-excited states at thermodynamic conditions that correspond to nonzero entropy density. The system is assumed to be isolated from and not interacting with any other system or environment.  The key question that is asked is whether or not this isolated many-body system ``thermalizes'': does it successfully serve as a thermal ``bath'' for itself and, under its own unitary quantum dynamics, bring all of its subsystems to thermal equilibrium with each other.  The MBL (localized) phase is the part of this system's phase diagram where the system remains (Anderson) localized near any nonthermal initial state, so it fails to bring itself to thermal equilibrium: it fails to thermalize.  Some recent reviews about MBL include \cite{dri,al,aabs,gp}.

Most uses of the strong-randomness RG for many-body quantum systems are for the study of the system's ground state and low-lying excited states.  In the study of MBL, on the other hand, the focus is on highly-excited states, often typical states that at thermal equilibrium would correspond to infinite temperature.  One main question that we want to address for MBL systems is whether or not the eigenstates of the system's dynamics, as well as its long-time dynamical states, are at thermal equilibrium.  The MBL phase is the regime where the system does not go to thermal equilibrium nor are the eigenstates of its dynamics at thermal equilibrium, even in the limits of large systems and long times.  It appears that a true MBL phase that remains localized in the standard thermodynamic limit and in the infinite-time limit is a possibility only for one-dimensional systems with short enough range interactions~\cite{drh,gh}, so those are the systems the remainder of this review considers.  

The dynamics of a MBL system may be due to a time-independent Hamiltonian, or due to a Floquet unitary operator that produces the dynamics for one period of a Hamiltonian that is periodic in time.    Many-body localization does not require randomness, for example it may occur due to nonrandom quasiperiodicity of the system~\cite{iyer}, but here we only consider MBL due to quenched randomness.  An example Hamiltonian is a Heisenberg spin-1/2 chain with a random field:
\begin{equation}
H = \sum_n [h_n S_{n,z}+ \vec{S}_n\cdot\vec{S}_{n+1}]~,
\end{equation}
with the quenched random fields $\{h_n\}$ at each site $n$ drawn independently and uniformly from $[-W,+W]$.  This is one of the most studied MBL models~\cite{ph}; it is sometimes called the ``standard model'' for MBL, but only because it is highly studied and not because it is the best choice for a model in which to study MBL.  The MBL phase is at large $W$ (strong random fields) for this model.  The precise location of the phase boundary is still not clear, but it appears to be at $W_c>15$.~\cite{mcklh,sels}

The MBL phase is a gapless quantum-critical dynamic phase, and the strong-randomness RG has been used within the MBL phase by various authors; much of this work is reviewed in~\cite{im2}.  Here I instead review some of the RG studies of the dynamic quantum phase transition between the MBL phase, where the system does not thermalize, and the thermal phase, where the system does thermalize in the long time and large system limits. Early numerical work on this MBL phase transition in one-dimensional systems found behavior consistent with exponential dynamical scaling (dynamic critical exponent $z\rightarrow\infty$), suggesting that it might be governed by an infinite-randomness fixed point~\cite{ph}.  Since then, there has been a series of publications steadily exploring and developing strong-randomness RG treatments of this MBL phase transition in one-dimensional systems with short-range interactions and/or hoppings~\cite{vha,pvp,zzdh,dvp,thmdr,gvs,dgpsv,mh,mhi,agv,skvg}. 

When one uses a strong-randomness RG to study a many-body quantum ground state, this is usually done working directly with the microscopic degrees of freedom of the system, and the highest-energy remaining local excitations are usually integrated out, leaving a renormalized system with fewer degrees of freedom, as in most renormalization groups.  For the MBL phase transition, on the other hand, a direct fully controlled RG calculation has not been found, so the RG is instead phenomenological and/or approximate, and in most cases both.  The number of degrees of freedom sets the density of states (many-body states per unit energy) and thus the thermodynamic entropy, which is a key ingredient in the physics of the MBL phase transition, so no true ``integrating out'' that removes degrees of freedom is done.  Thus the RG renormalizes (coarse-grains) the dynamics, scaling to longer times and larger length scales, but without the usual reduction of the number of degrees of freedom.  As a result, the RG in most cases does not work directly with the microscopic degrees of freedom, but instead with more coarse-grained measures of the local dynamics.

In strong-randomness RG treatments of the MBL phase transition, the one-dimensional system is to some extent coarse-grained in to ``blocks'', which are segments of the chain with certain coarse-grained properties~\cite{vha,pvp,zzdh,dvp,thmdr,gvs,dgpsv,mh,mhi,agv,skvg}.  In many of these works, the blocks are assumed to be of only two types: locally thermalizing blocks that are locally less random and/or more interacting, so these thermal blocks do become strongly entangled; and locally MBL blocks that are locally more random and/or less interacting, so these locally MBL blocks have localized states that remain area-law entangled unless thermalized by a nearby thermal block.  Such a ``binary'' classification of the local properties is only approximate when applied at small length scales.  However, since the resulting RG does exhibit a flow to infinite randomness in the MBL phase and at the phase transition, this binary description then becomes self-consistent and thus potentially asymptotically correct.  Multiple assumptions are used in developing these RGs, so there certainly remains the possibility that some important and relevant physics has been left out.

Near the fixed point governing the MBL phase transition, the locally thermalizing blocks are rare, occupying an asymptotically vanishing fraction of the length of the one-dimensional system as one renormalizes to the limit of low energy scales.  These widely-spaced thermal blocks are randomly placed, and their lengths have a probability distribution.  This strong-randomness RG is, as usual, a {\it functional} RG, with this distribution of thermal block lengths the function that is being renormalized.  

Each thermal block is locally thermalizing the nearby locally MBL regions.  As we renormalize to lower energy scales in this RG the locally thermal blocks thermalize the spins in the adjacent MBL blocks out to a distance that grows logarithmically with the (inverse) energy scale.  There are two ``events'' that happen as this RG flow runs: (i) A single thermal block may reach the limit of how many spins in the nearby MBL regions it can thermalize.  Once we renormalize below that energy scale, that thermal block is found to instead be localized (no longer able to spread more thermalization), so it is removed and that region is instead localized at this lower energy scale.  (ii) If two nearby thermal blocks manage to thermalize the entire MBL region that lies in between them, then they merge with each other and with this intermediate region that they have thermalized to make a new, much longer renormalized thermal block~\cite{mh,mhi}.  The latter process broadens the distribution of the lengths of the thermal blocks.  The energy scale of a thermal block is set by its many-body energy-level spacing which is exponentially small in its length, so this length distribution becoming broader means a flow to infinite randomness.

A nice analogy with the Kosterlitz-Thouless (KT) RG and transition has been found and developed~\cite{thmdr,gvs,dgpsv,mh,mhi}: In Kosterlitz-Thouless, if vortices and antivortices are forbidden, then the two-dimensional superfluid remains stable.  This critical KT superfluid phase is governed by a RG fixed line, with the superfluid stiffness divided by the temperature being the dimensionless parameter that varies along this fixed line.  The KT phase transition occurs where the RG flow near this fixed line changes from being stable to being unstable to allowing vortices and antivortices~\cite{kt,kosterlitz}. 

For the MBL transition in one-dimensional systems with quenched randomness, the role of vortices and antivortices is instead played by locally thermalizing rare regions (thermal blocks)
~\cite{dri,drh,mh,mhi}.  In this approach, the MBL phase is governed by a fixed line of the RG, with the dimensionless parameter that varies along the fixed line being the ratio of two lengths: one being the length per bit of thermodynamic entropy, and the other length being a decay length for the exponential dependence of the relaxation rate of spins on their distance from the nearest thermal block~\cite{drh,mh,mhi}.  The MBL phase is governed by the part of this fixed line that is stable to allowing locally thermalizing rare regions (thermal blocks); in the MBL phase these rare regions are RG-irrelevant, so the RG flow goes to this fixed line.  In the MBL phase these thermal blocks only manage to locally thermalize a limited number of spins in the adjacent MBL blocks before the energy scale is reached where this thermalization stops and the thermal block becomes instead localized.  The MBL-to-thermal phase transition is then governed by the point on this fixed line where these locally thermalizing rare regions (added thermal blocks) become RG-relevant, so the fixed line becomes unstable in the thermal phase and the RG flow heads off towards thermalization.  This occurs when, under the RG flow, enough of the thermal blocks can merge and produce much longer thermal blocks, so the fraction of the system occupied by the thermal blocks instead increases under the RG flow.

Although there is this strong {\it qualitative} analogy to the Kosterlitz-Thouless (KT) RG, the strong-randomness RG flow for the MBL transition is mathematically distinct~\cite{mhi}.  For KT, the RG is a two-parameter flow, with the vortex fugacity and the reduced superfluid stiffness being the two parameters.  For MBL, on the other hand, it is instead a {\it functional} RG, with the function being the probability distribution of the lengths of the locally thermalizing rare regions.  When two of these rare regions (thermal blocks) are close enough together to thermalize the locally MBL typical region in between them, then, under the RG flow, they merge in to a much longer rare region, extending the distribution function to those longer lengths~\cite{mh,mhi}.  This produces a rather direct connection within the functional RG flow between two very different length scales, a feature that is not present in the simpler Kosterlitz-Thouless RG flow.

Recent numerical work to quantitatively estimate where this asymptotic MBL-to-thermal phase transition occurs in the phase diagram of microscopically-defined spin-chain models has shown that it is actually very deep in the regime where the behavior of those models appears to be strongly localized for sample sizes and times accessible to standard numerical and experimental methods~\cite{mcklh,sels}.  Thus it currently appears that the physics of the asymptotic MBL phase transition as found in the strong-randomness RG may only apply on rather large length scales and thus only for extremely low energy scales (equivalently, extremely long times).  One way to put a ``positive spin'' on our present understanding of this situation is to note that the strong-randomness RG method allows one to develop what appears to be a controlled asymptotic low-energy theory of this phase transition, even though other approaches to studying this novel and theoretically challenging phase transition have not been able to reach those very low energy scales.

\section{conclusion}

The strong-randomness renormalization group methods are versions of the renormalization group (RG) that are useful for treating certain systems with quenched randomness, particularly those whose low-energy behavior is governed by an infinite-randomness fixed point.  These methods have wide applicability in certain one-dimensional quantum systems, as well as in various other systems, as reviewed in Refs.~\cite{im,im2,ar}.  More recently, such strong-randomness RG methods have been generalized to study the dynamic phase transition between many-body localization (MBL) and thermalization, as I have briefly reviewed here.

\section{acknowledgement}

I thank Michael Fisher for being a wonderful teacher, graduate adviser, and collaborator, and for all that he taught me about critical phenomena, the renormalization group, and many other things.  I am deeply indebted to him for his contributions to my education and development as a scientist.

\end{document}